\documentclass[12pt]{article}

\usepackage[super,sort&compress,comma]{natbib}

\usepackage{fancyhdr} 

\usepackage[section]{placeins}

\makeatletter \renewcommand\@biblabel[1]{$^{#1}$} \makeatother
\newcommand{\cen}[1]{\begin{center} #1 \end{center}}

\usepackage{hyperref}
\hypersetup{
    colorlinks,
    citecolor=blue,
    filecolor=blue,
    linkcolor=blue,
    urlcolor=blue
}

\usepackage[all]{hypcap}    

\usepackage{amsmath,mathtools,xfrac}
\usepackage{algorithm,algpseudocode}
\usepackage{optidef}
\def\argmin{\mathop{\rm arg\,min}}

\usepackage{amsfonts,amssymb,bbm}

\usepackage{graphicx,float}

\usepackage{enumerate,lscape,tabularx,booktabs}

\begin{document}

\cen{\sf {\Large {\bfseries A hyperparameter-tuning approach \\ to automated inverse planning} \\  
\vspace*{10mm}
K. Maass, A. Aravkin, M. Kim} \\
University of Washington, Seattle WA
}

\pagenumbering{roman}
\setcounter{page}{1}
\pagestyle{plain}

\begin{abstract}

\noindent {\bf Background:}
In current practice, radiotherapy inverse planning often requires treatment planners to modify multiple parameters in the treatment planning system's objective function to produce clinically acceptable plans.
Due to the manual steps in this process, plan quality can vary depending on the planning time available and the planner's skills.

\noindent {\bf Purpose:} 
This study investigates the feasibility of two hyperparameter-tuning methods for automated inverse planning.
Because this framework does not train a model on previously-optimized plans, it can be readily adapted to practice pattern changes, and the resulting plan quality is not limited by that of a training cohort.

\noindent{\bf Method:}
We retrospectively selected 10 patients who received lung SBRT using manually-generated clinical plans. 
We implemented random sampling (RS) and Bayesian optimization (BO) to automatically tune objective function parameters using linear--quadratic utility functions based on 11 clinical goals. 
Normalizing all plans to have PTV D95 equal to 48 Gy, we compared plan quality for the automatically-generated plans to the manually-generated plans.
We also investigated the impact of iteration count on the automatically-generated plans, comparing planning time and plan utility for RS and BO plans with and without stopping criteria.

\noindent{\bf Results:}
Without stopping criteria, the median planning time was 1.9 and 2.3 hours for RS and BO plans, respectively.
The organ-at-risk (OAR) doses in the RS and BO plans had a median percent difference (MPD) of 48.7\% and 60.4\% below clinical dose limits and an MPD of 2.8\% and 3.3\% below clinical plan doses.
With stopping criteria, the utility decreased by an MPD of 5.3\% and 3.9\% for RS and BO plans, but the median planning time was reduced to 0.5 and 0.7 hours, and the OAR doses still had an MPD of 42.9\% and 49.7\% below clinical dose limits and an MPD of 0.3\% and 1.8\% below clinical plan doses.

\noindent{\bf Conclusions:}
This study demonstrates that hyperparameter-tuning approaches to automated inverse planning can reduce the treatment planner's active planning time with plan quality that is similar to or better than manually-generated plans. 
\end{abstract}

\setlength{\baselineskip}{0.7cm} 

\pagenumbering{arabic}
\setcounter{page}{1}
\pagestyle{fancy}
\newpage

\section{Introduction}
\label{sec:intro}

In inverse planning for radiotherapy, the treatment planner translates a patient's clinical goals into a mathematical objective function, which is then minimized to determine the optimal treatment plan.
The objective function typically contains terms corresponding to different clinical goals, such as maximum doses or dose--volume objectives, and each term is assigned a weight factor to specify the relative importance of treating or sparing particular structures. 
Unfortunately, the treatment plan that minimizes a given objective function may not be clinically acceptable or optimal, either because clinical goals have not been met, or because further improvement can still be made, either with respect to better planning target volume (PTV) coverage or lower doses for organs-at-risk (OAR). 
To find a suitable plan, treatment planners often modify the objective function parameters through an inefficient trial-and-error approach, where the influence of individual parameters on the resulting treatment plan is not always intuitive and can depend on a variety of factors including model formulation, optimization algorithm, and patient anatomy.
This means that plan quality is often determined by the skills of an individual planner and the time available to try different parameter configurations \cite{bohsung2005imrt, chung2008can}.

To improve planning efficiency and plan quality, many efforts have been made to either guide or automate this ``manual planning" step.
One approach is to explore the relationships between the objective function parameters and the resulting dose distributions by means of an effect--volume histogram (EVH) \cite{alber2002toolsI} or an analytical sensitivity analysis \cite{alber2002toolsII, krause2008role, sobotta2008tools}.
In the first case, the EVH can identify organ subvolumes where weight adjustment can have the greatest impact, detect conflicting objectives, and help determine if goals can be met by either changing weights or including additional objective terms.
In the second case, a sensitivity analysis of the objective and constraint parameters can quantify the influence of each parameter with respect to the objective function, which can then be used to predict changes to the resulting plan without re-optimization.
This requires knowledge of the specific form of the objective and its Lagrangian function, which may not be available for all treatment planning systems.
When these functions are not known, an empirical sensitivity analysis can be performed by computing treatment plans for randomly-sampled objective function parameters \cite{lu2007reduced}.
This sensitivity information can then be used to classify parameters into three sets: a fixed set (often normalized to meet clinical specifications), an insensitive set, and a sensitive set, which allows planners to reduce the dimensionality of the problem by restricting their search to the latter.

Another approach, often referred to as knowledge-based planning (KBP), uses a library of previously-optimized plans to predict an optimal plan for a new patient.
In general, these approaches can predict dose--volume histograms (DVH), specific dose metrics, voxel-level doses, objective function weights, beam-related parameters, or quality assurance metrics using either one or more similar reference cases (atlas-based methods) or a predictive model trained on many cases (model-based methods) \cite{ge2019knowledge}.
For example, the dose distributions and DVHs from reference plans can be used to guide the optimization of new plans via the constraint that OAR doses cannot be worse than the reference \cite{fredriksson2012automated}.
In addition, correlations between the spatial relationships amongst OARs and PTVs, such as what is captured in the overlap--volume histogram or distance-to-target histogram, and the dose delivered in prior plans can be leveraged to develop models to predict expected OAR DVHs from patient geometry.
This can in turn be used to generate planning goals, optimization objectives, or objective function weights for new patients \cite{wu2011data, ziemer2017fully, babier2018inverse} or to develop quality-control tools that can automatically detect suboptimal plans \cite{appenzoller2012predicting}.
Features of patient anatomy can also be incorporated into models such as $k$-nearest neighbors or multinomial logistic regression to predict objective function weights to warm start the manual-planning process \cite{boutilier2015models}.
While these tools can reduce planning time by identifying a good starting point for a particular patient, or by indicating when an acceptable plan has been found, the resulting plan quality can be dependent upon the quality of the training set \cite{landers2018fully}.
Furthermore, there are only a limited number of studies that incorporate KBP methods into a fully-automated pipeline that generates a complete treatment plan \cite{mcintosh2017fully, babier2020knowledge, momin2021knowledge, visak2021automated}, so the planner may still need to manually tune plans by trial and error.
In smaller clinics, data sets of high-quality plans with similar patient geometry and objectives may not be readily available, necessitating predictive models validated across different institutions.
KBP approaches may also require training a new model when a practice pattern (clinical objective) changes, for example, a prescription dose escalation requiring a steeper dose gradient between PTVs and OARs.

To overcome the challenges of KBP, we propose a hyperparameter-tuning approach to automated inverse planning, which does not require a training set of high-quality previously-optimized plans.
Using a treatment plan utility function or other clinical criteria to quantitatively compare different treatment plans, the hyperparameter-tuning approach iteratively selects new objective function parameters to improve plan quality using various search methods.
Our approach uses a commercial treatment planning system (TPS) with scripting capability \cite{Raystation, Eclipse} as a black-box subroutine, facilitating integration with current practice.
Framing objective function parameters within the TPS as tunable hyperparameters also allows us to take advantage of recent advances in hyperparameter optimization from the machine learning (ML) community.
As opposed to KBP approaches, where the OAR dose metrics largely depend on those in the training cohort, or previous efforts to automatically tune the weights of clinical objectives \cite{wang2017development, landers2018fully, shen2019intelligent}, where the emphasis is only on meeting clinical goals, this approach also attempts to find lower OAR doses whenever feasible.

In the ML setting, hyperparameters can include values related to the structure of a learning model (e.g., regularization weights, number of features, or the architecture of a neural network) or the optimization approach used to train the model (e.g., learning rate, mini-batch size, or number of epochs).
This approach seeks to minimize a loss function with respect to its hyperparameters, which can be challenging due the the fact that the function is typically nonconvex, expensive to evaluate, and does not have a closed-form expression or access to its derivatives \cite{wang2013bayesian}. 
In the same way that treatment plan quality often depends on the experience of the planner, the manual tuning of ML hyperparameters can be considered a ``black art" requiring expert intuition and heuristics \cite{bergstra2012random, snoek2012practical}.
To improve efficiency, results, and reproducibility, ML researchers have developed a variety of approaches for automated hyperparameter optimization.
For a small number of hyperparameters, a naive grid search can be effective, but it suffers the curse of dimensionality and is thus infeasible for most problems \cite{bergstra2012random, hazan2017hyperparameter}.
In some cases, when certain continuity and differentiability conditions are satisfied, gradients with respect to continuous hyperparameters can be calculated using back-propagation and automatic differentiation, which can allow for the optimization of a large number of hyperparameters \cite{bengio2000gradient, maclaurin2015gradient}.
Two important classes of hyperparameter optimization methods include model-free and model-based approaches.
Some examples of model-free approaches include evolutionary \cite{hansen2001completely, li2019generalized} and random searches \cite{bergstra2012random, li2016hyperband}, which are amenable to parallel computing.
In Sequential Model-based Optimization (SMBO), the loss function is approximated by a surrogate that is cheaper to evaluate, and new hyperparameters are chosen by optimizing an acquisition function based on the surrogate.
Examples include Bayesian optimization using Gaussian processes \cite{bergstra2011algorithms, snoek2012practical} or random forests \cite{hutter2011sequential, thornton2013auto}.

In this paper, we explore the feasibility of using hyperparameter tuning approaches to automate inverse planning processes for patients with peripherally-located lung tumors treated with stereotactic body radiation therapy (SBRT) using the RayStation TPS \cite{Raystation}.
Specifically, we automatically tune objective function parameters using two different search methods, uniform random sampling and Bayesian optimization, implemented in the Python package Scikit-Optimize 0.8.1 \cite{head2020scikit}.
We compare our automatically-generated plans for 10 patients with the clinical plans manually generated by certified medical dosimetrists.
Using intuitive utility functions that can be personalized to reflect individual patient goals, we demonstrate that it is often possible to decrease doses to OARs below clinical goal limits, resulting in treatment plans that are as good as or better than manually-generated plans.
Furthermore, the automated approach requires less active planning time than the manual trial-and-error approach.

\section{Materials and Methods}

In the inverse-planning workflow, clinical goals are specified by the physician, often in terms of minimum or maximum doses, dose--volume objectives, average doses, or uniform doses.
It is then the job of the treatment planner to translate these goals into a mathematical objective function, which is optimized by a TPS to determine an acceptable dose distribution for the patient.
The TPS generally calculates beam parameters $x$ by solving a problem of the form
\begin{mini}
    {x }{f(x; w, \theta) := \sum_{i=1}^k w_i f_i(x; \theta_i),}{}{\label{eq:fmo}}
\end{mini}
where constituent functions $f_i$ penalize beam configurations that violate a particular clinical goal.
Positive weight factors $w_i$ are used to specify the relative importance of treating or sparing particular structures, while nonnegative dose parameters $\theta_i$ are chosen based on goal values $\gamma_i$.
Some constituent function types, such as those corresponding to dose--volume objectives, require additional parameters related to volume limits.
Examples of constituent functions formulated with quadratic penalties are given in Table~\ref{tab:constFuncs}, where $d_j: \mathbb{R}_{\geq0}^m \to \mathbb{R}_{\geq0}^{n_j}$ maps the $m$ beamlet intensity values to the $n_j$ voxel doses for the $j$th region of interest (ROI).
Depending upon the TPS, the specific penalties used for each type of constituent function may not be available to the planner.

\begin{table}[tpb]
    \centering
    \caption{Examples of constituent functions using quadratic penalties.}
    \begin{tabular}{l l l}
        \toprule
        ROI & Clinical Goal & Constituent Function \\
        \hline
        PTV$_j$ & Uniform dose $= \gamma_i$ & $\| d_j(x) - \theta_i \|_2^2$ \\
        PTV$_j$ & Minimum dose $\geq \gamma_i$ & $\| \min\{0, d_j(x) - \theta_i\} \|_2^2$ \\
        OAR$_j$ & Maximum dose $\leq \gamma_i$ & $\| \max\{0, d_j(x) - \theta_i\} \|_2^2$ \\
        \bottomrule
    \end{tabular} \vspace*{4ex}
    \label{tab:constFuncs}
\end{table}

\subsection{Hyperparameter tuning --- random sampling and Bayesian optimization}

For a fixed set of $k$ constituent functions, let $w \in \mathbb{R}_{>0}^k$ denote the vector of weight factors and $\theta \in \mathbb{R}_{\geq0}^k$ the vector of dose parameters.
Viewing the results of the TPS as a black-box function, we let $x: \mathbb{R}_{>0}^k \times \mathbb{R}_{\geq0}^k \to \mathbb{R}_{\geq0}^m$ return the solution to \eqref{eq:fmo} for a given $w$ and $\theta$.
Defining the treatment plan utility function $g: \mathbb{R}_{\geq0}^m \to \mathbb{R}$ to quantify treatment plan quality with respect to the clinical goal reference doses encoded in the vector $\gamma \in \mathbb{R}_{\geq0}^k$, our hyperparameter optimization problem can be expressed as
\begin{maxi!}[2]
    {w>0, \theta\geq0}{g[x(w,\theta); \gamma]}{\label{eq:hyperOpt}}{}
    \addConstraint{x(w,\theta) \in \argmin_{x\geq0}f(x; w, \theta).}
\end{maxi!}
We implemented two different hyperparameter tuning approaches, random sampling and Bayesian optimization, to solve \eqref{eq:hyperOpt}.
We also investigated the impact of iteration count by creating plans both with and without stopping criteria.

\paragraph{Random sampling.} 

For iterations $t = 1, \dots, T$, parameters $w^{(t)}$ and $\theta^{(t)}$ are sampled uniformly from a given range of values, then the TPS calculates the beams $x$ which determine the plan utility $g$. 
The optimal treatment plan is then determined by the parameters that produce the maximum utility amongst all sampled parameter vectors, as described in Algorithm~\ref{algo:RS}.
Sampling and evaluation can be done sequentially, or if the TPS allows, in parallel.

\paragraph{Bayesian optimization.}

Rather than sampling parameters $w^{(t)}$ and $\theta^{(t)}$ at random, the Bayesian approach approximates $g$ with a surrogate model, then determines which points to sample (i.e., which parameters to use next in the TPS objective function) by optimizing an acquisition function that quantifies the benefit of sampling from different locations based on 


\begin{algorithm}[H]
    \caption{Random sampling for $g(w,\theta)$ in~\eqref{eq:hyperOpt}.}
    \label{algo:RS}
    \begin{algorithmic}
        \State{Specify number of iterations $T$.}
        \For{$t = 1,\dots,T$}
            \State{Determine $(w^{(t)}, \theta^{(t)})$ via uniform random sampling.}
            \State{Calculate $g[x(w^{(t)}, \theta^{(t)}); \gamma]$ by solving \eqref{eq:fmo} with TPS.}
        \EndFor
        \State \Return{$\max_{\ell = 1,\dots,T}\{g[x(w^{(\ell)}, \theta^{(\ell)}); \gamma]\}$}
    \end{algorithmic}
\end{algorithm}

\vspace{4ex}
\begin{algorithm}[H]
    \caption{Bayesian optimization for $g(w,\theta)$ in~\eqref{eq:hyperOpt}.}
    \label{algo:BO}
    \begin{algorithmic}
        \State{Specify surrogate model, acquisition function, and number of iterations $T$.}
        \For{$t = 1,\dots,T$}
            \State{Optimize acquisition function to determine $(w^{(t)}, \theta^{(t)})$.}
            \State{Sample $g[x(w^{(t)}, \theta^{(t)}); \gamma]$ by solving \eqref{eq:fmo} with TPS.}
            \State{Update surrogate model using data $\{(w^{(\ell)}, \theta^{(\ell)}), g[x(w^{(\ell)}, \theta^{(\ell)}); \gamma]\}_{\ell=1,\dots,t}$.}
        \EndFor
        \State \Return{$\max_{\ell = 1,\dots,T}\{g[x(w^{(\ell)}, \theta^{(\ell)}); \gamma]\}$}
    \end{algorithmic}
\end{algorithm}

\noindent the values of the surrogate model.
The surrogate model is then refined after each iteration using a Bayesian posterior update, as described in Algorithm~\ref{algo:BO}.
Our implementation approximates the utility function $g$ using the nonparametric Gaussian process surrogate model.
A Gaussian process, defined in terms of mean and covariance functions, is a generalization of a multivariate Gaussian probability distribution, which is specified in terms of a mean vector and covariance matrix.
For the treatment planning problem, our Gaussian process represents a collection of random treatment plan utility variables $g(w, \theta)$, any finite subset of which are jointly Gaussian distributed.
In particular, for any input $(w, \theta)$, there is an associated random variable $g(w, \theta)$ with mean $\mu(w, \theta)$ and standard deviation $\sigma(w, \theta)$.
For simplicity, the mean function if often initially assumed to be zero, and the covariance is defined by a covariance function or kernel $k[(w, \theta), (w', \theta')]$, which specifies assumptions about the similarity between nearby points.

Our preliminary experiments revealed that ROI doses were more sensitive with respect to the dose parameters $\theta$ than the weight factors $w$, which motivated us to tune the dose parameters exclusively in the following comparative treatment-planning study.

\paragraph{Stopping criteria.}

To investigate the impact of iteration count on the automatically generated plans, in terms of both plan quality and planning time, we created plans using a set $T = 100$ iterations and using stopping criteria. 
Specifically, after a mandatory 20 iterations (where the first 20 iterations of the Bayesian routine correspond to the first 20 random samples), we terminated Algorithms~\ref{algo:RS} and \ref{algo:BO} if there was less than a 1\% increase within the 15 best utility values at iteration $t \leq T$.

\subsection{Comparative treatment-planning study}

We retrospectively selected 10 patients with peripherally-located lung tumors treated with SBRT.
For all cases, the objective function in \eqref{eq:fmo} consisted of 11 constituent functions corresponding to the clinical goals given in Table~\ref{tab:goals}, chosen according to RTOG 0915 \cite{videtic2015nrg}.
The MaxDVH type has the goal that the specified volume of an ROI does not exceed $\gamma_i$ in dose.
The D2cm is defined as a region that is at least 2 cm away from the tumor, and therefore the maximum dose in the D2cm indicates how quickly the dose falls off outside the PTV.
All weight factors were fixed at $w_i = 1$, and all dose parameters $\theta_i$ were tuned over the range $[\sfrac{\gamma_i}{4}, \gamma_i]$, with the exception of the PTV parameters.
Specifically, the PTV D95 parameter was fixed at $\theta_6 = \gamma_6$, and because the PTV maximum-dose parameter $\theta_7$ cannot be less than the PTV D95 parameter, it was instead tuned over the range $[\sfrac{(\gamma_7 + 3\gamma_6)}{4}, \gamma_7]$.
All treatment plans, both automatically and manually generated, were normalized so that exactly 95\% of the PTV received at least 48 Gy.

\begin{table}[tpb]
    \centering
    \caption{Clinical goals used in comparison study.}
    \begin{tabular}{l l l l l}
        \toprule
        $i$ & ROI & Type & Dose ($\gamma_i$) & Volume \\
        \hline
        1 & Chest Wall & MaxDVH & 30 Gy & 30 cm$^3$ \\
        2 & D2cm & MaxDose & 24--30 Gy$^*$ \\
        3 & Esophagus & MaxDVH  & 18.8 Gy & 5 cm$^3$ \\
        4 & Lungs & MaxDVH & 11.6 Gy & 1500 cm$^3$ \\
        5 & Lungs & MaxDVH & 12.4 Gy & 1000 cm$^3$ \\
        6 & PTV & MinDVH & 48 Gy & 95\% \\
        7 & PTV & MaxDose & 80 Gy \\
        8 & Ribs & MaxDVH & 32 Gy & 1 cm$^3$ \\
        9 & Ribs & MaxDose & 40 Gy \\
        10 & Spinal Cord & MaxDVH & 13.6 Gy & 1.2 cm$^3$ \\
        11 & Spinal Cord & MaxDose & 26 Gy \\
        \bottomrule
        &\scriptsize{$^*$PTV volume dependent} \vspace{5ex}
    \end{tabular}
    \label{tab:goals}
\end{table}

\begin{figure}
    \centering
    \hspace*{4em}
    \includegraphics[scale=0.7]{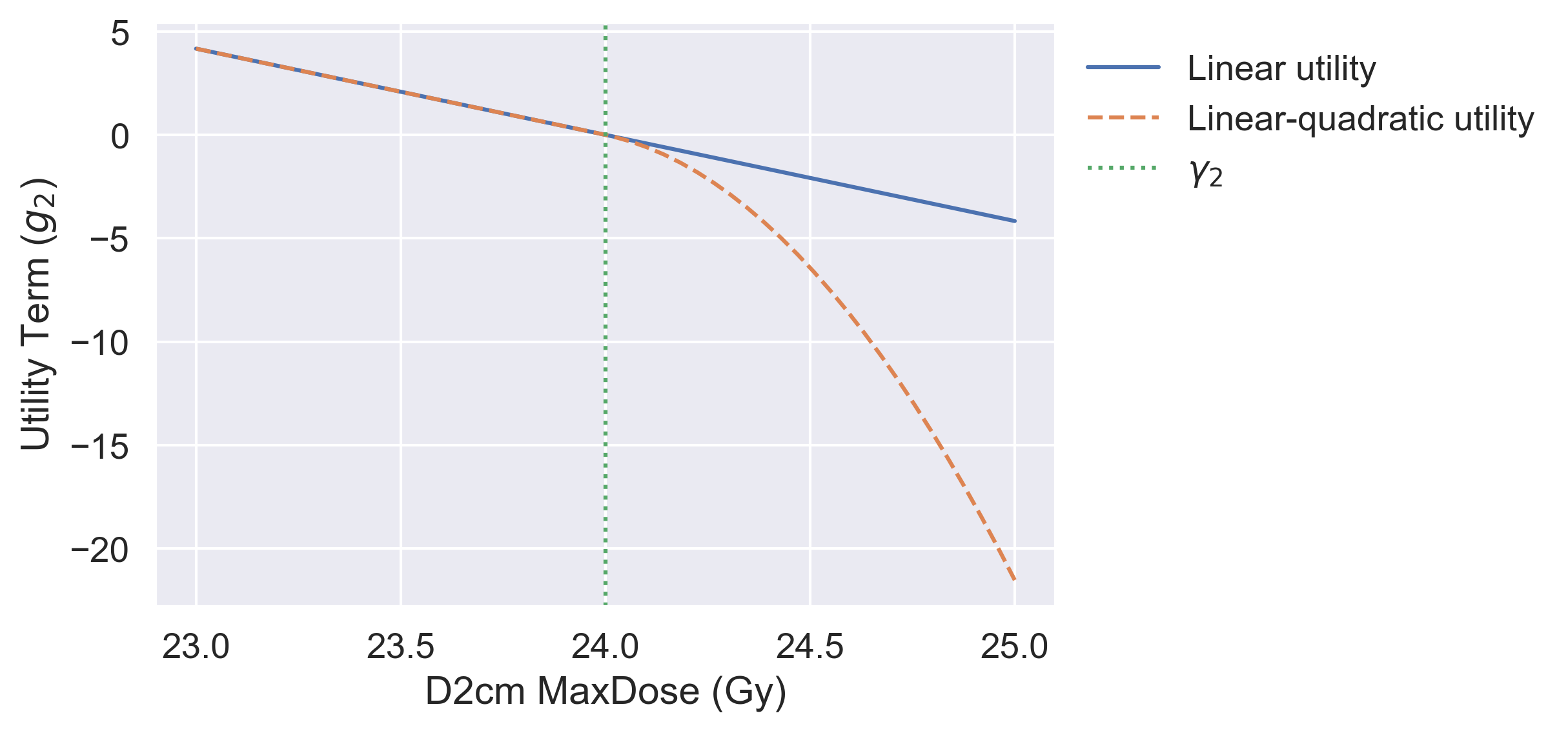}
    \caption{Linear and linear--quadratic utility terms corresponding to a maximum-dose clinical goal of 24 Gy for the D2cm ROI. \\}
    \label{fig:util}
\end{figure}

Our treatment plan utility function in \eqref{eq:hyperOpt} was defined as a linear combination of terms corresponding to each clinical goal,
\begin{equation}
    g[x(w, \theta); \gamma] := \sum_{i=1}^{11} g_i[x(w,\theta); \gamma_i],
    \label{eq:util}
\end{equation}
though the term corresponding to the PTV D95 could be omitted due to normalization. We used a combination of linear and linear--quadratic utility terms to reward plans that decrease doses below (or increase doses above) the clinical goal limits.
Our linear utility terms for maximum-dose objectives,
\begin{equation}
    g_i^{\ell}[x(w,\theta); \gamma_i] = \frac{100\{\gamma_i - d_j^{\max}[x(w,\theta)]\}}{\gamma_i},
\end{equation}
where $d_j^{\text{max}}$ calculates the maximum dose delivered to the $j$th ROI, can be interpreted as the percent decrease relative to the reference dose value, depicted as the solid blue line in Figure~\ref{fig:util}.
While the linear utility terms have an intuitive interpretation, they do not distinguish between doses that meet a clinical goal from doses that do not.
Specifically, a decrease to doses above $\gamma_i$ is rewarded the same as an equal decrease to doses below $\gamma_i$ (i.e., the slope of the utility is the same both above and below $\gamma_i$).
For clinical goals with higher priority, we instead used a linear--quadratic utility,
\begin{equation}
    g_i^{\ell q}[x(w, \theta); \gamma_i] = \begin{cases} g_i^{\ell}[x(w, \theta); \gamma_i] &\quad d_j^{\max}[x(w, \theta)] \leq \gamma_i, \\
    \{1 - g_i^{\ell}[x(w, \theta); \gamma_i]\} g_i^{\ell}[x(w, \theta); \gamma_i] & \quad d_j^{\max}[x(w, \theta)] > \gamma_i, \end{cases}
\end{equation}
depicted as the dashed orange line in Figure~\ref{fig:util}.
We used the linear--quadratic utility for all ROIs except for the chest wall and ribs, where we used the linear utility.
This was based on the clinical observation that the violation of the chest wall and rib clinical goals is often acceptable in practice.

Our approach was implemented with RayStation 8B, which provides scripting capabilities in CPython 3.6.5.
A call to our utility function updates the RayStation objective function dose parameters, runs RayStation's optimization routine, retrieves the resulting ROI doses, and returns the calculated utility value.
Our hyperparameter optimization problem was solved using functions {\tt dummy\_minimize} (random sampling) and {\tt gp\_minimize} (Bayesian optimization with Gaussian processes) from the 
SMBO package Scikit-Optimize 0.8.1, which is built upon Python packages NumPy, SciPy, and Scikit-Learn \cite{harris2020array, 2020SciPy-NMeth, scikit-learn}.
All of our code, data, and figures are openly available under the CC-BY license \cite{maass2021figures}.
We used the default settings for {\tt gp\_minimize}, including a Mat\'{e}rn kernel with automatically-tuned hyperparameters (e.g., length scale, covariance amplitude, and additive Gaussian noise), and the acquisition function {\tt gp\_hedge}, which randomly selects either the lower confidence bound, expected improvement, or probability of improvement acquisition function each iteration.

We compared the results of the automatically-generated plans, created with both random sampling and Bayesian optimization based on Gaussian processes, and the manually-generated clinically plans.
For the two hyperparameter tuning approaches, both with and without stopping criteria, we present results in terms of planning time, utility values, and the dosimetric differences between the optimal plans and both the clinical goals and the manually-generated clinical plans.

\section{Results}

We first present the results of the plans computed using random search (RS) and Bayesian optimization (BO) with $T = 100$ iterations (i.e., no stopping criteria).
We chose 100 iterations because our preliminary experiments showed empirically that this was sufficient time for the utility to converge in most cases.

\paragraph{Dosimetric comparison with clinical plans.}

\begin{figure}
    \centering
    \includegraphics[scale=0.45]{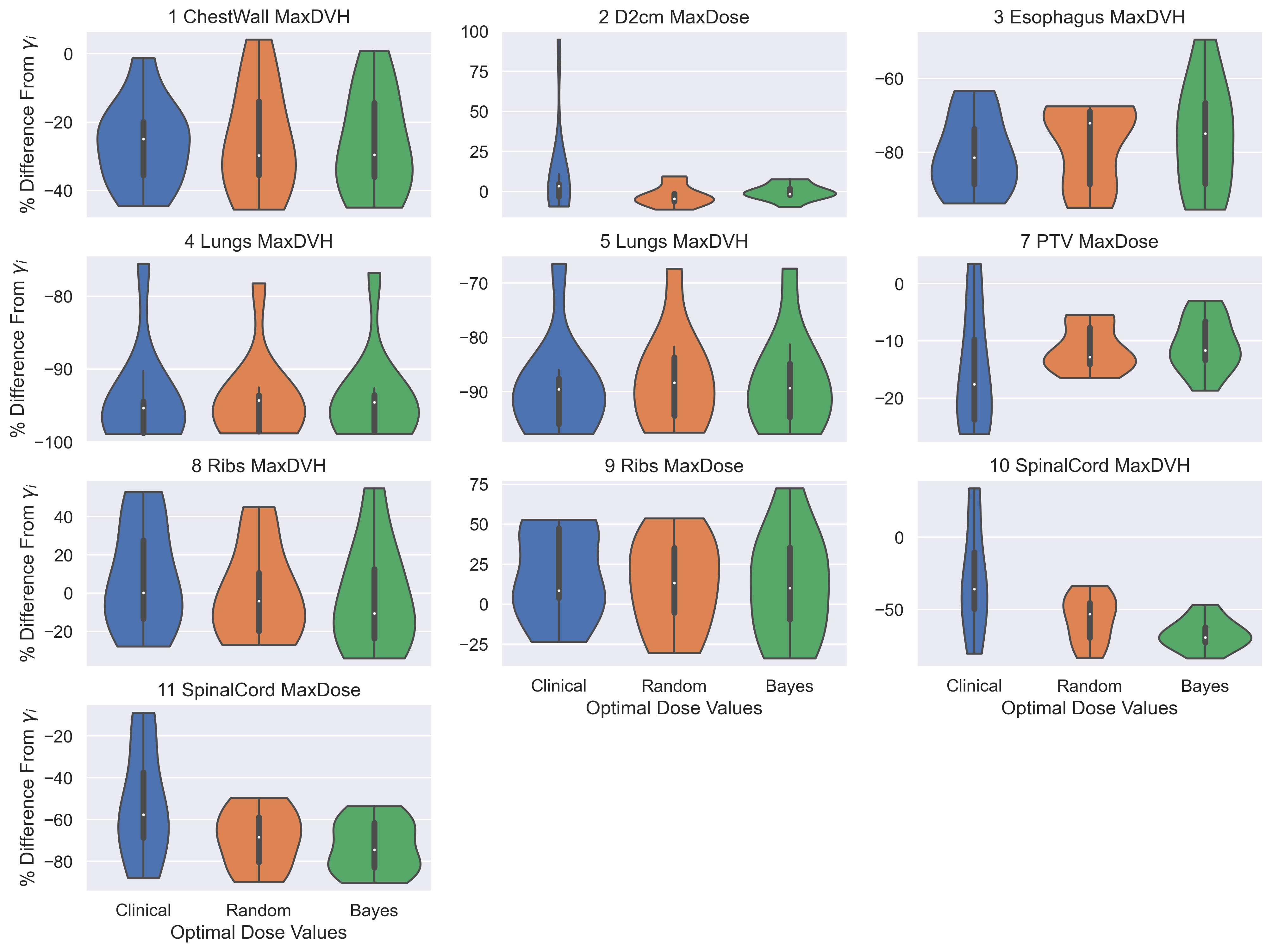}
    \caption{Plan doses corresponding to 10 clinical goals (PTV D95 doses excluded due to normalization), expressed as percent difference from clinical goal values $\gamma_i$, for manually-generated clinical plans (left), and plans automatically generated using random sampling (middle) and Bayesian optimization (right). \\}
    \label{fig:dose_full}
\end{figure}

For OAR doses, the RS and BO plans had a median percent difference (MPD) of 48.7\% and 60.4\% below the clinical goal values $\gamma_i$, respectively, with inner quartile ranges (IQR, i.e., Q1--Q3) of 7.52\%--83.4\% and 9.93\%--83.9\% below $\gamma_i$.
Furthermore, the OAR doses for the automatically-generated plans had an MPD of 2.82\% and 3.28\% below the doses in the manually-generated clinical plans, with an IQR of 14.8\% below--9.89\% above clinical plan doses for RS plans and an IQR of 18.9\% below--8.35\% above clinical plan doses for BO plans.
In Figure~\ref{fig:dose_full}, we present the ROI doses for the clinical (left), RS (middle), and BO (right) plans expressed as the percent difference from their clinical goal values.
Each violin includes both a box-and-whisker plot, indicating the data quartiles, and a kernel density estimation of the underlying distribution.

\paragraph{Optimal dose parameters.}

\begin{figure}
    \centering
    \includegraphics[scale=0.45]{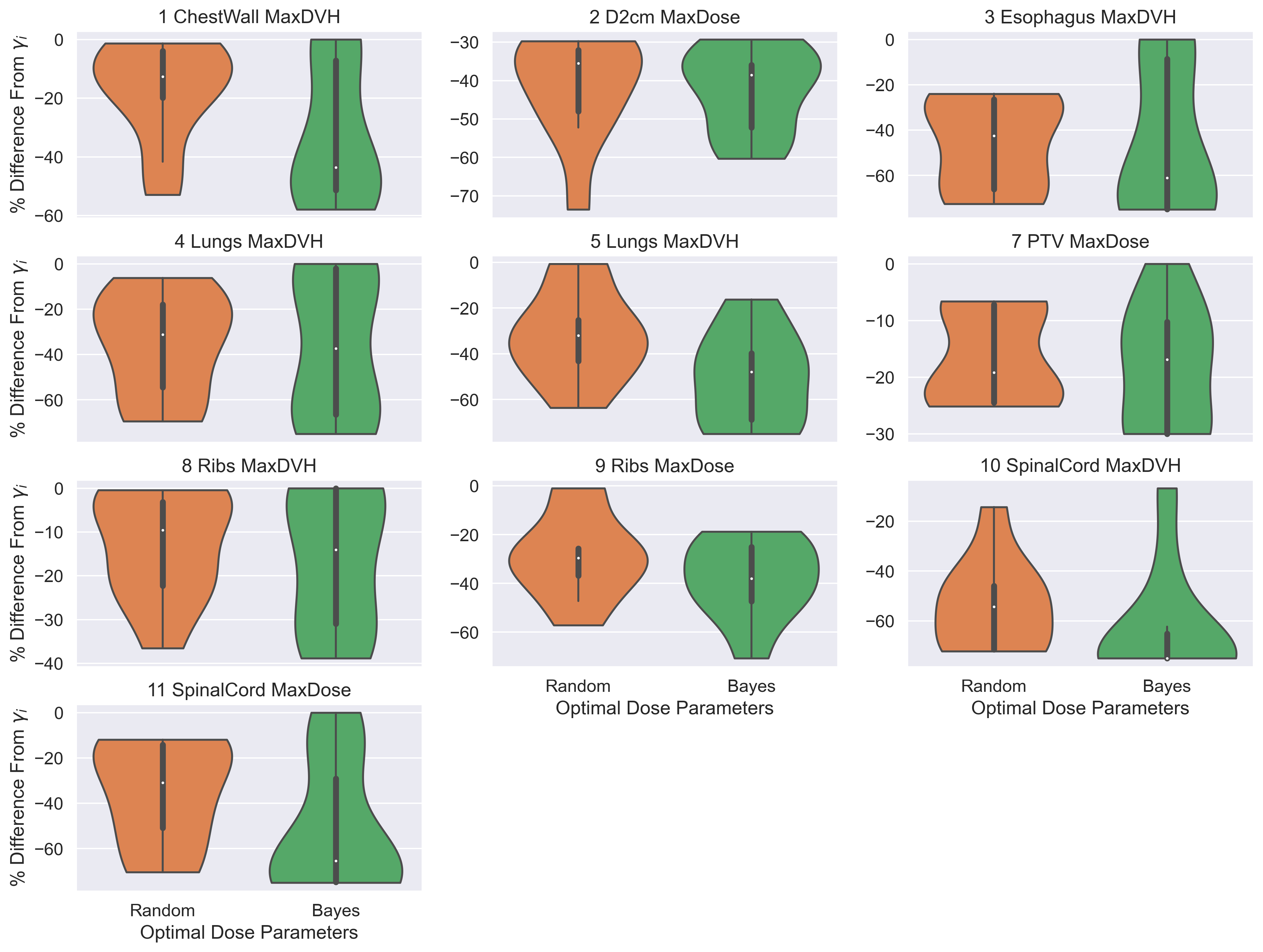}
    \caption{Optimal dose parameters $\theta_i$ for plans automatically generated using random sampling (left) and Bayesian optimization (right), expressed as percent difference from clinical goal values $\gamma_i$. \\}
    \label{fig:dosePars}
\end{figure}

In Figure~\ref{fig:dosePars}, we include the distribution of optimal dose parameters $\theta_i$, expressed as the percent difference from $\gamma_i$, for the automatically-generated plans.
The optimal dose parameters had an MPD of 29.7\% (IQR 14.5\%--47.4\%) and 38.2\% (IQR 16.9\%--63.8\%) below $\gamma_i$ for RS and BO plans, respectively.
We note that some optimal parameters never occurred in the upper portion of their respective ranges (e.g., D2cm MaxDose), while others never occurred in the lower portion (e.g., ribs MaxDVH). 
This may in part be due to the choice of linear vs. linear--quadratic utility terms.

\paragraph{Computation time.}

The random search method had a median computation time of 1.93 hours for 100 iterations, with an IQR of 1.56--2.41 hours.
This was on average 21.9 minutes faster than the Bayesian routine, which had a median computation time of 2.27 hours (IQR 2.06--2.53) for 100 iterations.

\paragraph{Utility.}

The plans computed with Bayesian optimization achieved the highest utility, with an MPD of 3.09\% (IQR 2.29\%--4.83\%) above the plans computed with random sampling and an MPD of 21.5\% (IQR 12.2\%--89.8\%) above the utilities of the clinical plans.

\paragraph{Effect of stopping criteria on planning time and plan quality.}

When the stopping criteria was applied, the treatment planning time was reduced to a median of 0.534 hours (IQR 0.445--0.610) for RS plans and a median of 0.673 hours (IQR 0.507--0.896) for BO plans.
The median reduction in planning time, compared to using 100 iterations, was 70.9\% (IQR 69.8\%--76.6\%) and 70.3\% (IQR 63.0\%--77.9\%) for the RS and BO plans, respectively. 
The median loss in utility due to stopping criteria was 5.31\% (IQR 3.15\%--14.6\%) and 3.87\% (IQR 1.45\%--9.92\%) for the RS and BO plans.
With stopping criteria, the BO plans still had the highest utility, with an MPD of 2.33\% (IQR 0.0736\%--20.3\%) above RS plan utilities and an MPD of 7.54\% (IQR 0.714\%--86.5\%) above the clinical plan utilities.

\begin{figure}
    \centering
    \includegraphics[scale=0.45]{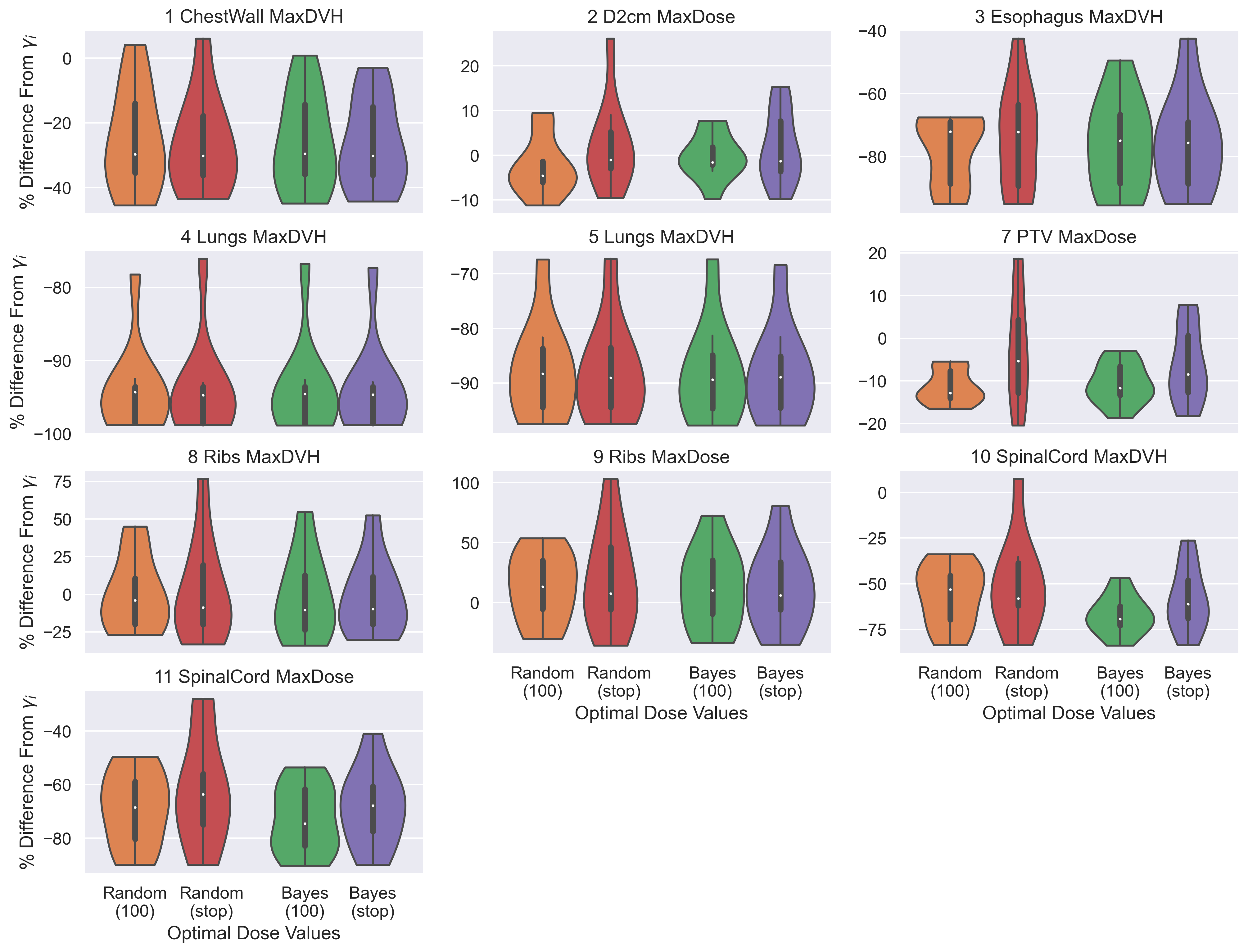}
    \caption{Plan doses corresponding to 10 clinical goals (PTV D95 doses excluded due to normalization), expressed as percent difference from clinical goal values $\gamma_i$, for plans automatically generated using random sampling (left) and Bayesian optimization (right), both with 100 iterations and with stopping criteria.}
    \label{fig:dose_stop}
\end{figure}

In Figure~\ref{fig:dose_stop}, we present the ROI doses for the automatically-generated plans, with 100 iterations and with stopping criteria, expressed as the percent difference from their clinical goal values $\gamma_i$.
The left two violins contain the distribution of RS plan doses, and the right two violins contain the BO plan counterparts.
The OAR doses for plans computed with stopping criteria had an MPD of 0.489\% and 0.722\% above the plans computed using 100 iterations, with an IQR of 2.25\% below--5.38\% above for RS plans and an IQR of 1.28\% below--5.35\% above for BO plans.
The OAR doses in the RS and BO plans computed with stopping criteria had an MPD of 42.9\% (IQR 5.75\%--83.2\%) and 49.7\% (IQR 8.01\%--83.5\%) below the clinical goal values, and an MPD of 0.254\% and 1.82\% below the doses in the manually-generated clinical plans, with an IQR of 9.88\% below--10.9\% above for RS plans and an IQR of 14.5\% below--7.23\% above for BO plans.

\section{Discussion}
Treatment planners create plans and evaluate plan quality based on clinical goals and experience.
Due to time constraints, the iterative process of manually adjusting optimization parameters often ends once an acceptable plan is identified, even if more improvements could be made.
KBP algorithms have been developed to improve planning efficiency and plan quality by automating the inverse planning process. 
However, most of these require creating a predictive model from a data set of prior clinical cases \cite{ge2019knowledge}, where plan quality depends on the plan quality in the training cohort.
A large data set of high-quality plans is not always available, and a change in clinical goals, goal values, or practice patterns may require a different predictive model.

To overcome these challenges, we proposed a hyperparameter-tuning approach to automated inverse planning, which directly optimizes the objective function parameters in the TPS instead of transferring prior knowledge to a new plan.
Because the objective function parameters are directly optimized, we can use any arbitrary clinical goals and goal values, which could be helpful when clinicians need to customize clinical goals to accommodate patient-specific concerns or adjust to standard practice pattern changes.
A hyperparameter-tuning approach can be particularly useful when it is unclear which parameters have the most influence on the resulting plan, for example, when working with unusual, patient-specific anatomy or with types of clinical objectives (e.g., biological objectives) that a planner has less experience with.
In these cases, tuning the weights of each clinical objective type may help planners to identify the most influential objective types and parameters.

Our approach can also explore a range of objective function parameters beyond the neighborhood of clinical goal values, with utility functions designed to minimize OAR doses whenever feasible. 
In fact, we observed that many of the automatically-generated plans were achieved with parameters on the lower boundary of their search ranges, as shown in Figure~\ref{fig:dosePars}, which may not be intuitive to human treatment planners.
For example, the optimal dose parameters for the spinal cord MaxDVH had an MPD of 54.3\% and 75.0\% below $\gamma_{10}$ for RS and BO plans, respectively, and the resulting doses had a MPD of 53.1\% and 69.3\% below $\gamma_{10}$ (the clinical plan doses had an MPD of 36.0\% below $\gamma_{10}$).
This supports the observation made in Holdsworth {\it et al.} \cite{holdsworth2012use} that more flexibility in the search space can lead to improved treatment plans.
We also note that manually-generated plans do not always include all clinical objectives in the TPS objective function if the planner expects that certain clinical goals are easily achievable, either from past experience or based on patient geometries (e.g., the OAR is very far from the PTV).
However, this can potentially miss the opportunity to lower the dose to an OAR when it is physically feasible.

The random search and Bayesian optimization routines we implemented had different advantages and disadvantages.
On the one hand, the RS plans required less computation time, both with and without stopping criteria, than the BO plans.
This is most likely due to the fact that the latter method also computed a Bayes posterior surrogate model and optimized an acquisition function every iteration.
On the other hand, the BO routines were able to learn about the sensitivity of treatment plan quality with respect to the objective function parameters, and thus restrict its search to better regions of the parameter space, resulting in higher utility and lower OAR doses than the RS plans.
Two other considerations are parallelization, which is more straightforward for randomized algorithms, and dimensionality.
Specifically, Bayesian optimization methods have been shown to work efficiently for problems of moderate dimension; however, their success is often limited to no more than 10--20 parameters \cite{wang2013bayesian, kandasamy2015high, maclaurin2015gradient}, at which point they may behave similarly to random methods.
Therefore, scaling the Bayesian approach to higher dimensions remains a challenge, and efforts to reduce problem dimension can lead to increased efficiency.

There are many potentials directions to expand upon the work in this paper.
For example, a more comprehensive study on effective stopping conditions could reduce treatment planning time while maintaining similar plan quality, and developing methods to automatically identify sensitive parameters and effective parameter ranges could guide dimensionality reduction to improve efficiency and utility. 
Other aspects of our implementation, including utility function design (e.g., using other function shapes such as sigmoids) and the choice of acquisition function and the surrogate model's covariance kernel, could also enhance performance.
Additionally, further advances in ML hyperparameter optimization can continue to motivate future work, including techniques such as early stopping \cite{li2016hyperband}, parallel implementations for Bayesian optimization \cite{snoek2012practical}, taking advantage of utility functions with additive structure \cite{kandasamy2015high}, or collaborative approaches that can tune multiple similar problems simultaneously \cite{bardenet2013collaborative}.

\section{Conclusions}
We demonstrated the feasibility and potential benefit of two hyperparameter tuning approaches to automated inverse planning through a comparative treatment-planning study on retrospective patient cases of peripherally-located lung tumors treated with SBRT.
Specifically, we automatically generated treatment plans for 10 patients using both random sampling and Bayesian optimization with Gaussian processes to automatically tune TPS objective function parameters using linear and linear--quadratic utility functions based on clinical goals.
Without stopping criteria, the median treatment planning time (no planner intervention required) was 1.93 hours (IQR 1.56--2.41) and 2.27 hours (IQR 2.06--2.53) for random sampling and Bayesian optimization, respectively, and plan quality was similar to or better than the manually-generated clinical plans.
With stopping criteria, these computation times were reduced to 0.534 hours (IQR 0.445--0.610) and 0.673 hours (IQR 0.507--0.896).
Our approach can be easily integrated with any commercial TPS with a scripting interface.
Furthermore, it does not require training a model on a library of previously-optimized plans, as in most KBP approaches, and therefore the resulting plan quality is not limited by that of a training cohort.
Consequently, our approach can be readily adapted to a practice pattern change, and it is able to lower doses to critical organs below clinical goal values whenever feasible.

\newpage 

\addcontentsline{toc}{section}{\numberline{}References}
\vspace*{-20mm}

\bibliography{references.bib} 


\bibliographystyle{./medphy.bst}


\end{document}